\documentclass[12pt]{article}

\usepackage{graphicx}  % Include figure files
\usepackage{amsmath}
\usepackage{amsfonts}
\usepackage{amssymb}

\usepackage{multirow}

\usepackage{color}

\begin{document}

  \title{Turing patterns with space varying diffusion coefficients:
    eigenfunctions satisfying the Legendre equation}

  \author{Elkinn A. Calder\'{o}n-Barreto, Jos\'{e} L. Arag\'{o}n\footnote{jlaragon@unam.mx} \\
    Centro de F\'{\i}sica Aplicada y Tecnolog\'{\i}a
    Avanzada, \\  Universidad Nacional Aut\'{o}noma de M\'{e}xico. \\
    Boulevard Juriquilla 3001 Juriquilla 76230 Quer\'{e}taro, Mexico.}

\maketitle
  
\begin{abstract}
The problem of pattern formation in reaction-diffusion systems with
space varying diffusion is studied. On this regard, it has already
been shown that the necessary conditions for Turing instability
are the same as for the case of homogeneous diffusion, but the
sufficient conditions depend on the specific space dependence of the
diffusion coefficient. In this work we consider the particular case
when the operator of the spectral Sturm-Liouville problem associated
with the general reaction-diffusion system has the Legendre
polynomials as eigenfunctions. We then take a step forward,
generalizing the standard weakly nonlinear analysis for these
eigenfunctions, instead of the eigenfunctions of the Laplace
operator. With the proposed generalization, conditions can be
established for the formation of stripped or spotted patterns, which
are verified numerically, and compared with the case of homogeneous
diffusion, using the Schnakenberg reaction diffusion system. Our
results enrich the field of pattern formation and the generalization
of the nonlinear analysis developed here can also be of interest in
other fields as well as motivate further generalization by using
general orthogonal functions.
\end{abstract}

% \linenumbers


\section{Introduction}\label{sec1}

In a ground-breaking work from 1952 Alan Turing laid the foundations
for chemical morphogenesis by proposing a reaction-diffusion theory for
pattern formation \cite{turing1952}. Turing showed that two or more
chemical substances, called morphogens, which react and diffuse in a
medium such as a tissue, can induce stable periodic patterns through
linear instability of a spatially uniform state. Turing's important
discovery is that, under certain conditions, a stable spatially
uniform state, in the absence of diffusion, can become unstable under
nonuniform perturbations (introduced as random initial conditions) due
to diffusion. This is the so-called Turing instability, which produces
a spatially non-homogeneous steady state, that is, a spatial
pattern. From a mathematical perspective, the Turing instability
mechanism has two components. First, it is necessary to derive the
necessary and sufficient conditions for diffusion-driven instability
of the spatially uniform state and the initiation of the spatial
pattern. This is achieved by a linear analysis, looking for a solution
that grows in time; this analysis results in the selection of a
dominant mode which is responsible for pattern initiation. Once the
instability criterion is stablished, the determined dominant mode
grows exponentially and is not valid at all times; and the long-time
evolution of the pattern, and the determination of the type of
pattern, should be studied by means of a nonlinear analysis
\cite{murray2003ii}.

The original Turing model became a master piece of mathematical
modeling in biology and the best known model for explaining biological
pattern formation. However, in recent years there have been great
advances in understanding both the mathematical and biological aspects
of this theory, including its generalization to a range of settings
beyond the assumptions established by Turing \cite{kondo,krause}.

One possible generalization of the original theory is to consider
spatial heterogeneity as a way to obtain irregular patterns. The
experimental evidence that in some biological systems spatial
inhomogeneities are importan to regulate patterns lead to several
generalizations of the reaction-diffusion problem: when one of the
diffusion coefficients either depends on the spatial variables
\cite{maini1992,benson1998} or is discontinuous
\cite{maini1992,wei2009}; time-dependent \cite{sherrat1995} or
concentration-dependent diffusion coefficients
\cite{roussel2004,gambino,das}; spatially varying parameters
\cite{page2003,page2005}; reaction-diffusion system in a channel with
the projected Fick-Jacobs-Zwanzig operator (with a diffusion
coefficient that depends on the longitudinal coordinate)
\cite{chacon}; and pattern formation with superdiffusion \cite{liu} or
anomalous diffusion \cite{hernandez,khudhair}. A recent advance in the
study of pattern formation in spatially heterogeneous
reaction-diffusion systems was made in \cite{vangorder}, where a more
general instability criterion, which can be applied to spatially
heterogeneous systems, was proposed as well as a procedure to calculate
nonlinear mode coefficients as a way to understand the influences of
each spatial mode on the long-time evolution of a pattern.

In this work, we consider the problem of a reaction-diffusion equation
when the diffusion coefficient depends explicitly on the space
variables,
$\nabla \cdot \left( \mathcal{D} ( {\mathbf x} ) \nabla {\mathbf u}
\right)$ for the particular case when the eigenfunctions of this
operator are the Legendre polynomials, $L_p(x)$ in 1D and
$L_{ij}(x,y) = L_i(x) L_j(x)$ in 2D. For this purpose, we used
eigenfunction expansions, as in \cite{dillon,kozak,vangorder}. Using
this approach, it has been shown that the linear stability properties
of the system are the same as those of a reaction-diffusion system
with constant diffusion coefficients \cite{vangorder}. The main
contribution of this paper lies in the nonlinear analysis, where we
propose a generalization of the standard weakly nonlinear analysis of
the reaction-diffusion system using Legendre polynomials. The
parameter regions for producing stripes or spots can be identified
from the developed general nonlinear analysis.

Numerical simulations using finite elements were performed using the
Schnakenberg system to verify the predictions of the proposed
nonlinear analysis and to provide examples of one-dimensional and
two-dimensional patterns, where the non-homogeneity of the final
patterns, clearly influenced for the properties of the Legendre
function, are visible.

The generalization of the standard nonlinear analysis using the
orthogonal functions associated with the Legendre equations developed
here can be of interest in the field of pattern formation and other
areas such as climate modeling, where some one-dimensional models
derived from the Budyko-Seller climate model contain a Legendre type
operator as the studied in this work \cite{north,hetzer}.

\section{Linear Turing analysis for spatially varying diffusion}
 \label{sec:linear}

 Because the linear stability analysis is the first step in the
 mathematical study of pattern formation, in this section, we follow
 the generalization of the classical Turing instability analysis
 developed in \cite{vangorder} for various scenarios of spatially
 heterogeneous systems, where spatially varying diffusion is a
 particular case. In this STUDY it is shown that the necessary
 conditions for Turing instability when the diffusion is homogeneous
 are the same as for the case of space varying diffusion coefficients,
 but the sufficient conditions may generally change. Henceforth, we
 will gather the results of this general theory applicable to our
 problem. This also allows us to establish essential ideas and
 notation.

The general non-dimensional form of a two-chemicals reaction-diffusion
system with a space varying diffusion coefficient, is:
\begin{equation}
  \label{eqn:general}
  \frac{\partial \mathbf{u}}{\partial t} = D \nabla \cdot \left(
    \mathcal{D}(\mathbf{x}) \nabla \mathbf{u} \right) + \eta F
  (\mathbf{u}). 
\end{equation}
where $\mathbf{u} = (u,v)$ is a vector of two species defined on a
bounded domain $\Omega \subset \mathbb{R}^n$, $D$ is the diagonal
matrix of diffusion coefficients $D_u$ and $D_v$ of $u$ and $v$,
respectively, $\mathcal{D}(\mathbf{x})$ is a function that describes
the spatial variation of the diffusion rate, $F = (f, g)^T$ is the
reaction kinetics, and $\eta$ is a non-dimensional coefficient related
to the size of the space domain. It is assumed that the system
(\ref{eqn:general}) is subject to Neumann (zero flux) boundary
conditions:
\begin{equation}
  \label{eqn:generalb}
  \mathbf{n} \cdot \left( \mathcal{D}( \mathbf{x} ) \nabla \mathbf{u}
  \right) = 0, \;\;\; \textrm{on} \; \mathbf{x} \in \partial \Omega .
\end{equation}
  
For simplicity, in what follows we assume $D_u=d$ and $D_v=1$, that
is, $D = \textrm{diag}[d,1]$. This is equivalent to scaling
(\ref{eqn:general}) and does not imply a loss of generality.
  
By following the standard procedure, if
$\textbf{u} = \textbf{u}_0=(u_0, v_0)$ is a steady state, that is
$\mathbf{f} (\mathbf{u}_0) = \mathbf{0}$, we consider a perturbation
of the form
$\textbf{u}(\textbf{x},t) = \textbf{u}_0 + \varepsilon
\textbf{v}(\textbf{x},t)$, where $0 < \varepsilon \ll 1$. The
perturbation $\textbf{v}(\textbf{x},t)$ is governed by the linear
problem \cite{vangorder}:
\begin{eqnarray}
  \label{eqn:generalin}
\frac{\partial \mathbf{v}}{\partial t} &=& D \nabla \cdot \left(
  \mathcal{D}(\mathbf{x}) \nabla \mathbf{v} \right) + \eta J \mathbf{v},
  \;\;\; \textrm{in} \; \Omega \nonumber \\
  \mathbf{n} \cdot \left( \mathcal{D}( \mathbf{x} ) \nabla \mathbf{v}
  \right) &=& 0, \;\;\; \textrm{on} \; \partial \Omega ,
\end{eqnarray}
where $J$ is the Jacobian matrix associated with the reaction kinetics
$F$, evaluated at the steady state $\textbf{u}_0$.

Consider the spectral problem
\begin{eqnarray}
  \label{eqn:spectral}
\nabla \cdot \left( \mathcal{D}(\mathbf{x}) \nabla \Phi \right)
  &=& -\rho \; \Phi,  \;\;\; \textrm{in} \; \Omega \nonumber \\
  \mathbf{n} \cdot \left( \mathcal{D}( \mathbf{x} ) \nabla \Phi
  \right) &=& 0, \;\;\; \textrm{on} \; \partial \Omega .
\end{eqnarray}
When solutions exist, there is an infinite but countable set of real
eigenvalues $\rho_i$, which can be ordered according to increasing
magnitude: $0 = \rho_0 < \rho_1 \leq \rho_2 \leq \cdots$, and the
corresponding set of eigenfunctions $\Phi_0, \Phi_1, \ldots$, form a
complete basis for functions on $\Omega$ satisfying the given boundary
conditions. Therefore, it seems reasonable to propose a solution to
(\ref{eqn:generalin}) as a linear combination of eigenfunctions:
\begin{equation}
  \label{eqn:eigenexp}
\mathbf{v}(\mathbf{x},t) = \sum_{n=0}^\infty \mathbf{C}_n (t) \Phi_n (
\mathbf{x}) ,
\end{equation}
where $\mathbf{C}_n \in \mathbb{R}^2$, $n=0, 1, 2, \ldots$.

Using all of the above, in \cite{vangorder} it is shown that the
evolution of the $k$th Turing mode coefficient vector $\mathbf{C}_k$
is given by
\begin{equation}
  \label{eqn:cequation}
  \frac{d \mathbf{C}_k}{d t} = \left( - \rho_m D + \eta J \right)
  \mathbf{C}_k , \;\;\; k = 0, 1, 2, \ldots ,
\end{equation}
which is the same system as that obtained for homogeneous diffusion,
with the only difference being the eigenvalues $\rho_m$ defined by the
specific choice of $\mathcal{D}(\mathbf{x})$.

Assuming $\mathbf{C}_k (t) = e^{\lambda_k t} \mathbf{c}_k$, for
$\mathbf{c}_k$ a constant vector and $\lambda_k \in \mathbb{C}$,the
values of $\lambda_k$ for a Turing instability are given by the
solution of the second order polynomial equation:
\begin{equation}
  \label{eqn:lambdas0}
 \lambda_k ^2 - \alpha(\rho_k) \lambda_k + \beta(\rho_k) = 0,
\end{equation}
where
\begin{equation}
  \label{eqn:alfabeta}
 \alpha(\rho_k) = \textrm{tr}(- \rho_k D + \eta J), \;\;\; \textrm{and}
 \;\;\; \beta(\rho_k) = \textrm{det}(- \rho_k D + \eta J) .
\end{equation}

% The solution of \ref{eqn:lambdas0} which gives an instability are
% \cite{vangorder}:
% \begin{equation}
%   \label{eqn:lambda}
%   \lambda_k = \frac{-(d+1) \rho_k + \textrm{tr}(J)}{2} + \frac{1}{2}
%   \sqrt{(1-d)^2 \rho_k ^2 + 2 (1-d) (J_{11}-J_{22}) \rho_k +
%     \textrm{tr}(J)^2 - \textrm{det} (J)}.
% \end{equation}

Linear stability is guaranteed if $\textrm{tr}(J) < 0$ and
$\textrm{det}(J) > 0$. For the steady state to be unstable to spatial
perturbations we require $\textrm{Re}(\lambda_k)>0$, which is
fulfilled if and only if $\beta(\rho_k) < 0$, which is the standard
condition for Turing instability in a spatially homogeneous system
\cite{vangorder}. The solutions of (\ref{eqn:lambdas0}) are evaluated
for each $k$ satisfying the above conditions to determine the largest
value of $\lambda_k$.

The critical value of $d$ and the critical value of $\rho$ are
obtained from the conditions $\beta (\rho_c) = 0$ and
$d \beta (\rho_c) / d \rho_c = 0$, respectively, yielding
\begin{equation}
  \label{eqn:dc}
d_c = \frac{ \left( J_{11}J_{22} - 2 J_{12} J_{21}  \right) - 2 \sqrt{-
    J_{12} J_{21} \textrm{det}(J)}}{J_{22}^2} ,
\end{equation}
and
\begin{equation}
\rho _c = \eta \frac{J_{11} + d_c J_{22}}{2 d_c} .
\label{criticoWaveNumber}
\end{equation}

% $ = \eta \left( \frac{\textrm{det}(J)}{d_c} \right)^{1/2}$

It is worth mentioning that although (\ref{criticoWaveNumber}) gives
the theoretical maximizing space eigenvalue $\rho_k$, this maximum may
not be attained since the wavenumbers $k$ form a discrete set. The
true maximizer will be the $\rho_k$ which makes the real part of
$\lambda_k$, the solution of (\ref{eqn:lambdas0}), maximal
\cite{vangorder}. On the other hand, the scale parameter $\eta$ plays
an important role because the range of wavenumbers that satisfy the
instability conditions depends on this parameter. Therefore $\eta$ can
be increased to ensure that allowable wavenumbers exist in the
unstable range of $k$ \cite{murray2003ii}. This is illustrated in
the following example.

\subsection{Example: the Legendre equation}
 \label{sec:example}

 As a particular example, consider the Schnakenberg reaction-diffusion
 system \cite{schnakenberg}, which has been frequently used because of
 its simple structure. In dimensionless form, we consider the
 following one-dimensional system in the space domain
 $\Omega = [-1,1]$:
\begin{subequations}
  \label{eqn:schnakenberg}
\begin{align}
  \frac{\partial u}{\partial t} & = d \;  \frac{\partial}{\partial x} \left(
                                  \mathcal{D}(x) \frac{\partial u}{\partial x}
                                  \right) + \eta \left( a - u + u^2 v
                                  \right) \\    
  \frac{\partial v}{\partial t}& =\; \; \frac{\partial}{\partial x} \left(
                                 \mathcal{D}(x) \frac{\partial v}{\partial x}
                                 \right) + \eta \left( b - u^2 v \right) \\
  \mathcal{D} (x) \frac{\partial u}{\partial x} &=0, \hspace{0.25cm}
                                                  \mathcal{D} (x)
                                                  \frac{\partial
                                                  v}{\partial x} =0
                                                  \hspace{0.25cm}
                                                  \textrm{at} \;\;
                                                  x=-1,1 , 
\end{align}
\end{subequations}
where $a$ and $b$ are positive constants. The Schnakenberg model is
the simplest known two-component chemical reaction system that allows
a limit cycle solution. Such a model involves three reactions (one of
which is autocatalytic) for two chemical products $X$ and $Y$ and two
chemical sources $A$ and $B$; in (\ref{eqn:schnakenberg}), $x$, $y$,
$a$, and $b$ are the respective concentrations. It has been one of the
most studied models because it is simple and well behaved, and it has
been applied to different problems of pattern formation, including
plant root hair initiation \cite{brena} and reaction-diffusion
equation systems coupled with convective flow \cite{wei}.

The spectral problem (\ref{eqn:spectral}) is the Sturm-Liouville
problem:
\begin{eqnarray}
  \label{eqn:spectralsl}
  \frac{d}{d x} \left( \mathcal{D}(x) \frac{d \Phi}{d x} \right)
  &=& -\rho \Phi,  \;\;\; \textrm{in} \; \Omega=[-1,1] \nonumber \\
  \mathcal{D}(x) \frac{d \Phi}{d x} &=& 0, \;\;\; \textrm{at} \;\; x=-1,1.
\end{eqnarray}

\begin{figure}[b!]
  \centering
  \begin{tabular}{c}
    \includegraphics[width=8cm]{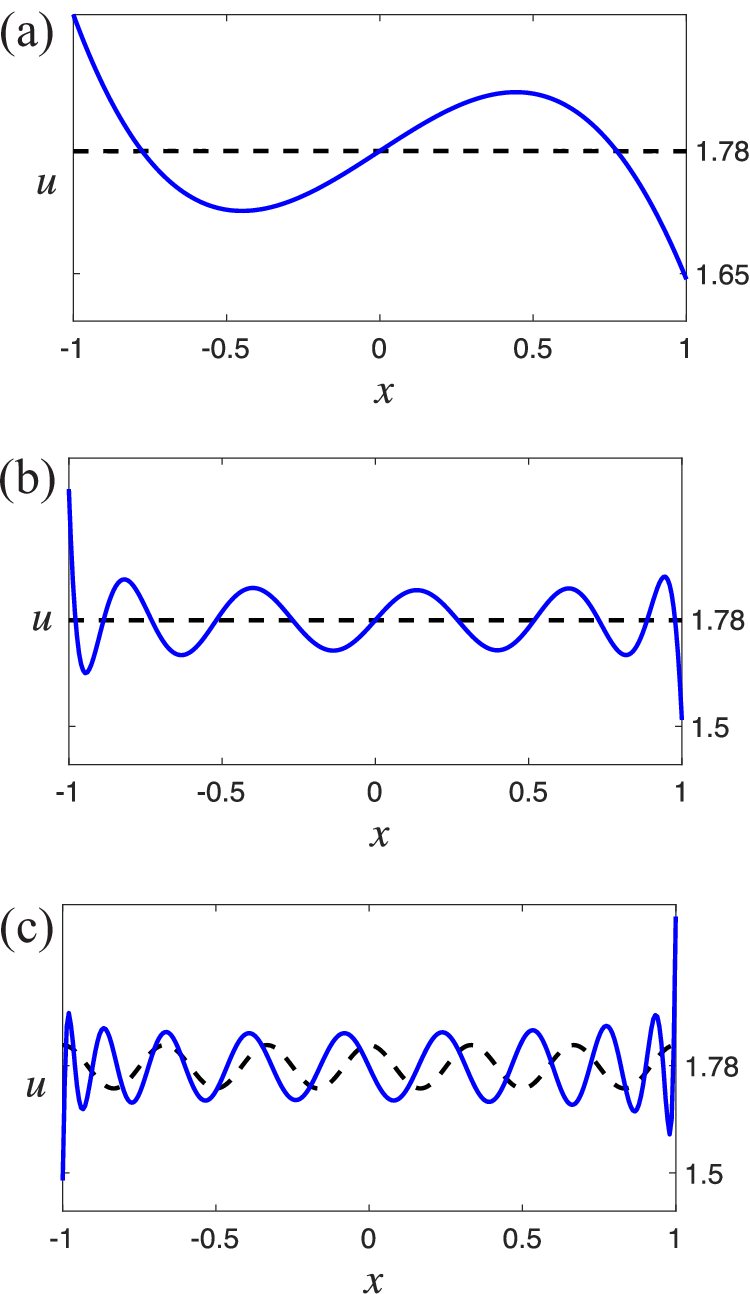}
  \end{tabular}
  \caption{The patterns obtained at $T=1000$ seconds with Legendre
    diffusion (blue continuous line) are compared tho those obtained
    with homogeneous diffusion (black dashed line).  \textbf{(a)}
    $\eta = 1.11571$, \textbf{(b)} $\eta = 12.2729$, and \textbf{(c)}
    $\eta = 35.3310$.}
  \label{fig:1Dcomparison}
\end{figure}

Let us consider first the case of homogeneous diffusion
$\mathcal{D}(x)=1$. In this case, we have $\Phi _k = \cos (k x)$ and
$\rho _k = k^2 = (n \pi)^2$. The system (\ref{eqn:schnakenberg})
has a single equilibrium state at $(u_0,v_0)=(a+b,b/(a+b)^2)$, and the
critical values (\ref{eqn:dc}) and (\ref{criticoWaveNumber}) become
\begin{equation}
  \label{eqn:dcS}
  d_c = \frac{(a+b) (a+3 b) - 2 \sqrt{2 b (a+b)^3}}{(a+b)^4},
\end{equation}
and
\begin{equation}
  \label{eqn:lcS}
  k_c ^2 = \frac{(a+b)^4 \eta}{\sqrt{2 b (a+b)^3} - (a+b)^2} . 
\end{equation}

Now, if $ \mathcal{D}(x) = 1-x^2$ the eigenfunctions of the
Sturm-Liouville problem (\ref{eqn:spectralsl}) $\Phi_k = P_k(x)$ are the
Legendre Polynomials of degree $k \in \mathbb{N}$, with eigenvalues
$\rho_k = k(k+1)$. In this case $d_c$ is also (\ref{eqn:dcS}), and
\begin{equation}
  \label{eqn:lcS2}
  k_c (k_c + 1) = \frac{(a+b)^4 \eta}{\sqrt{2 b (a+b)^3} - (a+b)^2} . 
\end{equation}

We numerically solve (\ref{eqn:schnakenberg}) with $a=0.289$ and
$b=1.49$, and with (\ref{eqn:dcS}) we calculate $d_c$ yielding
$d_c =0.02735$. For Legendre diffusion, $k$ is an integer; therefore
we choose $k_c = 3, 11, 19$ and from (\ref{eqn:lcS2}) the
corresponding value of the scale parameter $\eta$ is calculated. This
yields $\eta =1.11571$, $\eta = 12.2729$ and $\eta = 35.3311$. Note
that for homogeneous diffusion, $k_c$ is given by (\ref{eqn:lcS}), for
which we have $k_c = 3.4641$, $k_c = 11.4891$ and $k_c = 19.4936$,
respectively. In Fig. \ref{fig:1Dcomparison}, patterns obtained at
$T=1000$ seconds with Legendre diffusion are compared with those
obtained with homogeneous diffusion for the values given above.

Note that for the first two values of $\eta$, no pattern with
homogeneous diffusion is formed however, for Legendre, diffusion
patterns with the expected spatial modes are produced. For
$\eta = 35.3310$ both patterns are formed with the expected spatial
mode and the changes in wavelength in the case of Legendre diffusion
is also observed.

For two-dimensional problems, the Schnakenberg reaction-diffusion
system:
\begin{subequations}
  \label{eqn:schnakenberg2d}
\begin{align}
  \frac{\partial u}{\partial t} & = d \;  \nabla \cdot \left(
                                  \mathcal{D}( {\mathbf x}) \nabla u
                                  \right) + \eta \left( a - u + u^2 v
                                  \right) , \\    
  \frac{\partial v}{\partial t}& =\; \; \nabla \cdot \left(
                                 \mathcal{D}( {\mathbf x}) \nabla v
                                 \right) + \eta \left( b - u^2 v
                                 \right) ,
\end{align}
\end{subequations}
is considered in the space domain $\Omega = [-1,1] \times [-1,1]$, and
\begin{equation}
  \label{eqn:boundaries}
\mathcal{D} ({\mathbf x} ) \nabla u \cdot {\mathbf
  n} =0, \;\;\; \textrm{and}  \;\;\; \mathcal{D} ({\mathbf x}) \nabla v \cdot
       {\mathbf n}=0 ,
\end{equation}
on the boundaries of the square $\Omega$.

For the spatial variation of the diffusion rate, $\mathcal{D} (x,y)$,
we propose the following anisotropic tensor expression:
\begin{equation}
\mathcal{D} (x,y) = \left(
\begin{array}{cc}
 \left(
\begin{array}{cc}
 1-x^2 & 0 \\
 0 & 1-y^2 \\
\end{array}
\right) & 0 \\
 0 & \left(
\begin{array}{cc}
 1-x^2 & 0 \\
 0 & 1-y^2 \\
\end{array}
\right) \\
\end{array}
\right) .
\label{SturmModel}
\end{equation} 

With this matrix, the spectral problem (\ref{eqn:spectral}), in the
space domain $\Omega = [-1,1] \times [-1,1]$, becomes
\begin{subequations}
  \label{SturmModel2}
\begin{align}
  \frac{\partial }{\partial x}\left(\left(1-x^2\right) \frac{\partial
  \Phi_1}{\partial x}\right)+\frac{\partial }{\partial
  y}\left(\left(1-y^2\right) \frac{\partial \Phi_1}{\partial y}\right)
  &= - \rho \Phi_1 , \\
  \frac{\partial }{\partial x}\left(\left(1-x^2\right) \frac{\partial
  \Phi_2}{\partial x}\right)+\frac{\partial }{\partial
  y}\left(\left(1-y^2\right) \frac{\partial \Phi_2}{\partial y}\right)
  &= - \rho \Phi_2 .
\end{align}
\end{subequations}

By separation of variables it can be shown that the eigenfunctions of
this problem are
\begin{equation}
  \Phi_m = \Phi_{ij}(x,y) = P_i (x) P_j(y),
\end{equation}
where $m=1,2$, and eigenvalues
\begin{equation}
  \label{eqn:lind2}
  \rho = k_{ij} = i (i+1) + j (j+1).
\end{equation}

\begin{figure}[t!]
  \centering
  \begin{tabular}{c}
    \includegraphics[width=10cm]{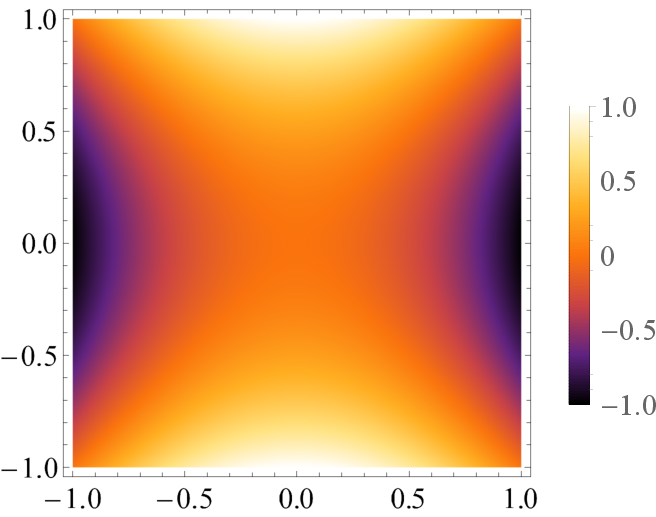}
  \end{tabular}
  \caption{A graphical interpretation of the diffusion produced by
    (\ref{SturmModel}); for each $(x,y)$, the difference in the
    diffusion coefficients, $(1-x^2) - (1-y^2)$, is plotted on a color
    scale. Positive values (light colors) indicate that concentrations
    move faster along the $x$-axis whereas negative values (dark
    colors) indicate that concentrations move faster along the
    $y$-axis.}
  \label{fig:figure2}
\end{figure}

The choice of (\ref{SturmModel}) is forced by mathematical simplicity
as any other choice could lead to a spectral problem that cannot be
solved, as is the case with a simpler choice, such as
$\mathcal{D}_{11} = \mathcal{D}_{22} = (1-x^2)(1-y^2)$. However, the
form (\ref{SturmModel}) is interesting in itself because, on the one
hand, it produces a highly anisotropic diffusion, as described in
Fig. \ref{fig:figure2}, where an interpretation of the spatial
variation of the diffusion is given; for each $(x,y)$, the difference
in the diffusion coefficients, $(1-x^2) - (1-y^2)$, is plotted on a
color scale. Therefore, positive values (light colors) indicate that
concentrations move faster along the $x$-axis whereas negative values
(dark colors) indicate that concentrations move faster along the $y$
axis. Thus, an initial concentration located near the central part of
the right (or left) edge, moves faster along the $y$-axis than along
the $x$-axis. In contrast, an initial concentration located near the
central part of the top (or bottom) edge, moves faster along the
$x$-axis. However, with (\ref{SturmModel}), the operator in
(\ref{SturmModel2}) is degenerate in the square
$\Omega = [-1,1] \times [-1,1]$. For instance, the wave number
$k_{i0}=k_{0i}$ in (\ref{eqn:lind2}) has associate the eigenfunctions
$\Phi_{i 0}$, $\Phi_{0 i}$, and the linear combination of both. This
was required to generate stripped and spotted patterns.

Because the linear analysis in the previous section is independent of
the spatial dimension, the critical values $d_c$ and $k_c$ are given
by (\ref{eqn:dcS}) and (\ref{eqn:lcS2}), respectively. Some
numerically obtained two-dimensional patterns will be shown and
explained in Sec. \ref{sec:numerical}.

All numerical simulations in this study were carried out with the
general-purposes finite element COMSOL
Multiphysics\textsuperscript{\tiny\textregistered} software
\cite{comsol}, using Lagrange quadratic elements. A time step of
$\Delta t = 0.001$ was used and the initial conditions were uniformly
randomly generated around the steady state, with a range of $0.1$. For
one-dimensional problems, a highly refined mesh consisting of 100
evenly spaced space nodes was considered, and for two-dimensional
problems, the refined mesh contained 24,912 domain elements and 400
boundary elements. Both spatial and time discretizations were tested by
starting with larger values and verifying that further decrements did
not modify the final form of the solution.

\section{Nonlinear analysis for spatially varying diffusion: The
  Legendre eigenfunctions case}

The spatially inhomogeneous solutions predicted by the linear analysis
grow exponentially and then are not valid at all times. By means of a
weakly nonlinear analysis, we can obtain approximate solutions that
are valid for all the time. The standard nonlinear analysis developed
for reaction-diffusion systems is based on the eigenfunctions of the
Laplace operator, that is, $\mathcal{D}(\mathbf{x}) = 1$ in
(\ref{eqn:spectral}). To consider the Legendre operator
(\ref{SturmModel}), it is necessary to generalize this analysis based
on the eigenfunctions $\Phi_{ij}(x,y) = P_i (x) P_j(y)$. Then the
methodology developed in References \cite{ermentrout,mahar,zhu} will
be generalized for the orthogonal functions $P_n$, and the conditions
for the generation of stripes and spots will be deduced. Before
proceeding, we present some general results.

The Taylor series in two variables $\mathbf{u} = (u,v)$ around
  $\mathbf{u}_0 = (u_0, v_0)$, for the field $F = \eta (f,g)^T$, is
  given by:
\begin{equation}
F (\mathbf{u}) = F_0 + \eta J \mathbf{u} + Q(\mathbf{u},
\mathbf{u}) + C(\mathbf{u}, \mathbf{u},\mathbf{u})+\cdots.
\end{equation}
Here $F_0= \textbf{0} $, and the linear term is the Jacobian matrix:
\begin{equation}
  \label{jacobiano}
J =
\left(\begin{array}{c c}
f_u & f_v\\
g_u & g_v
      \end{array}\right)_{(u_0, v_0)} .  
\end{equation}
The general expressions for the quadratic and cubic terms, which will
be required later, are

\begin{eqnarray}
Q( \mathbf{u}_1, \mathbf{u}_2) &=& \frac{\eta}{2} \left(\begin{array}{c}
f_{uu} u_{1} u_{2} + f_{uv} u_{1} v_{2} + f_{vu} v_{1} u_{2} + f_{vv} v_{1} v_{2}\\
g_{uu} u_{1} u_{2} + g_{uv} u_{1} v_{2} + g_{vu} v_{1} u_{2} + g_{vv} v_{1} v_{2}
\end{array}\right)_{(u_0, v_0)} ,
\label{terminocuadratico} \\
  C( \mathbf{u}, \mathbf{u}, \mathbf{u}) &=& \frac{\eta}{6}\left(
\begin{array}{c}
  f_{u^3} u_{1}^3 + 3 f_{u^2v} u_{1}^2 v_{1} + 3 f_{uv^2}  u_{1}
  v_{1}^2 + f_{v^3}  v_{1}^3\\
  g_{u^3} u_{1}^3 + 3 g_{u^2v} u_{1}^2 v_{1} + 3 g_{uv^2} u_{1}
  v_{1}^2 + g_{v^3} v_{1}^3
\end{array}\right)_{(u_0, v_0)} ,
\label{terminocubico}
\end{eqnarray}
where $\mathbf{u} = (u, v)$, $\mathbf{u}_1 = (u_1, v_1)$, and
$\mathbf{u}_2 = (u_2, v_2)$.

Finally, let $F_1 = (f_1, g_1)$ and $F_2 = (f_2, g_2)$ be two vector
fields in a domain $\Omega$. The inner product of $F_1$ and $F_2$ is
defined by
\begin{equation}
  \label{eqn:ip}
\left\langle F_1 | F_2 \right\rangle =   \left\langle (f_1, g_1) \cdot
  (f_2, g_2) \right\rangle = \int_\Omega (f_1 f_2 + g_1 g_2)
d\mathbf{x} .
\end{equation}

\subsection{Multiscale method}

Consider the following form of the reaction-diffusion system:
\begin{equation}
  \label{eqn:RD}
  \frac{\partial {\mathbf u}}{\partial t} =  D \nabla \cdot
  (\mathcal{D}(\mathbf{x}) \nabla {\mathbf u}) + F ( {\mathbf u}; p )
  , \;\;\; \textrm{in}  \; \Omega
\end{equation}
where
$F ( {\mathbf u}; p ) = \eta (f ( {\mathbf u}; p ), g ( {\mathbf u}; p
))^{\textrm{T}}$, $p$ is a parameter of the system, and $\eta$ is a
scale factor. The domain of the system is
$\Omega = [-1,1] \times [-1,1]$, and Neumann (zero-flux) boundary
conditions are considered:
\begin{equation}
  \label{eqn:BC}
  \mathbf{n} \cdot \left( \mathcal{D}( \mathbf{x} ) \nabla \mathbf{u}
  \right) = 0, \;\;\; \textrm{on} \; \partial \Omega .
\end{equation}

The wavelength $\lambda$ is perturbed around its bifurcation value
$\lambda_c$ as $\lambda=\lambda_c+\epsilon^2$, and close to the
bifurcation point it is assumed that the solution of the equation is
\begin{equation}
   \label{SE}
   \mathbf{u} = \mathbf{u}_0 + \hat{\mathbf{u}} = \mathbf{u}_0 +
   \left( \epsilon \mathbf{u_1}+\epsilon^2 \mathbf{u_2}+\epsilon^3 
     \mathbf{u_3} + \cdots \right).
\end{equation}
The case $ 0 <\epsilon < 1$ is considered and
a slow time scale is introduced:
\begin{equation}
T=\hat{w} t ,
\label{SS}
\end{equation}
where
\begin{equation}
\hat{w} =\epsilon w_1+\epsilon^2 w_2+\cdots  .
\label{SW}
\end{equation}  

A Taylor series expansion of $F$ in (\ref{eqn:RD}), around the
stable equilibrium point $\mathbf{u}_0$,  produces
\begin{equation}
F(\mathbf{u})= \eta J \hat{\mathbf{u}} +
Q (\hat{\mathbf{u}},\hat{\mathbf{u}}) +
C (\hat{ \mathbf{u}}, \hat{ \mathbf{u}},\hat{
  \mathbf{u}})+\cdots ,
\label{eq:TF}
\end{equation}
where $\eta J$, $Q$ and $C$ are, respectively,
the lineal quadratic and cubic terms.

By substituting (\ref{SE}) and (\ref{SS}) into (\ref{eqn:RD}), one
obtains
\begin{equation}
  \hat{w} \frac{\partial \hat{ \mathbf{u}}}{\partial T} = D
  \nabla \cdot \mathcal{D}(\mathbf{x}) \nabla \hat{ \mathbf{u}} +
  \eta J \hat{ \mathbf{u}} + 
  Q (\hat{ \mathbf{u}},\hat{ \mathbf{u}}) +
  C (\hat{ \mathbf{u}}, \hat{ \mathbf{u}}, \hat{ \mathbf{u}})+\cdots ,
\label{E}
\end{equation}

If additionally the parameter $p$ of the model is perturbed about its
value in a critical set:
\[
p = p_c + \hat{p} = p_c + (\epsilon p_1 + \epsilon ^2 p_2 + \cdots),
\]
then the expansion of (\ref{E}) in a Taylor series around $p_c$,
produces

\begin{eqnarray}
  \label{E2}
  \hat{w} \frac{\partial \hat{ \mathbf{u}}}{\partial T} &=&  (D
      \nabla \cdot \mathcal{D}(\mathbf{x}) \nabla)^c \hat{ \mathbf{u}} +
      \eta J^c \hat{ \mathbf{u}} + Q^c(\hat{\mathbf{u}},
      \hat{ \mathbf{u}}) + C^c(\hat{\mathbf{u}}, \hat{\mathbf{u}},
      \hat{ \mathbf{u}})+\cdots + \nonumber \\
      & & \hat{p}((D \nabla \cdot \mathcal{D}(\mathbf{x})
  \nabla)_p^c \hat{\mathbf{u}} + \eta J_p^c \hat{
  \mathbf{u}} + Q_p^c(\hat{ \mathbf{u}},\hat{ \mathbf{u}}) +
          \cdots) ,
\end{eqnarray}
where the subscripts stand for derivative with respect to $p$ and the
superscripts mean the evaluation at $p_c + \hat{p}$.

By proposing a solution of the form (\ref{SE}), this perturbation method
produces a hierarchy of linear differential equations:

\begin{align}
  \mathtt{O}(\epsilon): \hspace{0.3cm}& \mathcal{L} \mathbf{u}_1=\mathbf{0} ,
  \label{sisMulScala1} \\ 
  \mathtt{O}(\epsilon^2): \hspace{0.3cm}& \mathcal{L} \mathbf{u}_2 =
                                          Q (\mathbf{u}_1,
                                          \mathbf{u}_1) + p_1 \eta J_p^c  \mathbf{u}_1 -w_1
                                          \frac{\partial
                                          \mathbf{u}_1}{\partial
                                          T} , \label{sisMulScala2}
  \\
  \mathtt{O}(\epsilon^3):  \hspace{0.3cm}& \mathcal{L} \mathbf{u}_3 = Q
                                           (\mathbf{u}_1,\mathbf{u}_2)
                                           + C (\mathbf{u}_1,
                                           \mathbf{u}_1, \mathbf{u}_1) + p_2 \eta J_p^c
                                           \mathbf{u}_1 + p_1 \eta  J_p^c  \mathbf{u}_2
                                           + \nonumber \\
                                      &\hspace{1.25cm} p_1\mathbf{Q}^c( \mathbf{u}_1, \mathbf{u}_1) -w_1
                                        \frac{\partial \mathbf{u}_2}{\partial T}-w_2
                                        \frac{\partial \mathbf{u}_1}{\partial T},   
\end{align}
where
$\mathcal{L} = (-\eta J - D \nabla \cdot
\mathcal{D}(\mathbf{x}) \nabla)$. These equations will be solved in
what follows.

\subsubsection*{$O(\epsilon)$ equations}
In the presence of a Turing instability
$\rho_{k_c}=k_{i_c 0} = k_{0 j_c}$ (see Eq.  \ref{eqn:lind2}), and the
solution of the linear problem can be written as the sum of $N$
spatial modes. We consider the special case when two spatial modes are
present ($N=2$):
\begin{equation}
 \label{solu1}
 \mathbf{u}_1 = \mathbf{V} ^{(1)} a(T) P_{i_c}(x) + \bar{\mathbf{V}}
 ^{(1)} \bar{a}(T) P_{j_c}(y) ,
\end{equation}
where $P_k$ is the Legendre polynomial of degree $k$, and $k_{i_c j_c}$
satisfies the diffusion-driven instability conditions.

From (\ref{sisMulScala1}) we have:
\begin{align}
  (-\eta J + k_{i_c j_c} D )\mathbf{V} ^{(1)}&=\mathbf{0} ,
\end{align}
and the same for $ \bar{\mathbf{V}} ^{(1)}$.  The solution of the
above equation can be obtained and chosen to be a unitary vector, this
is:
\begin{equation}
  \mathbf{V} ^{(1)} = \frac{1}{\sqrt{(-\eta \; g_v + d \; k_{i_c j_c})^2 + \eta^2
      g_u^2}}  \left(
    \begin{array}{c}
      -\eta \; g_v + d \; k_{i_c j_c} \\
      \eta \; g_u
    \end{array}
  \right) .
\end{equation}
Since $i_c = j_c$, then $\mathbf{V} ^{(1)} = \bar{\mathbf{V}}
^{(1)}$. 

In what follows, to simplify notation, we assume that
$P_{i_c} = P_{i_c} (x)$, $\bar{P}_{j_c} = P_{j_c} (y)$, $a = a(T)$, and
$\bar{a} = \bar{a}(T)$.

\subsubsection*{$O(\epsilon^2)$ equations}
To solve the second equation, we begin by applying the Fredholm Alternative:

\begin{align}
\label{FA1}
\left\langle \mathbf{u}^* \middle\vert\ \mathbf{Q}(\mathbf{u}_1,
\mathbf{u}_1)+p_1 \eta J_p^c  \mathbf{u}_1 -w_1
\frac{\partial
	\mathbf{u}_1}{\partial
	T} \right\rangle  =0 ,
\end{align}
where $\mathbf{u}^*$ is in the null-space of the adjoint. Since the
solution of the adjoint problem has the same form as the solution of
the linear case, we propose
\begin{equation}
  \label{eqn:uast1}
\mathbf{u}^* = \mathbf{V}^{*} a P_{i_c}, 
\end{equation}
or
\begin{equation}
  \label{eqn:uast2}
  \mathbf{u}^*=\bar{\mathbf{V}}^{*}\bar{a} \bar{P}_{j_c}. 
\end{equation}

The explicit form of $Q (\mathbf{u}_1, \mathbf{u}_1)$ can be obtained
from (\ref{terminocuadratico}) using (\ref{solu1}). By doing this, we
get:
\begin{eqnarray}
\label{QU1U1}  
  \mathbf{Q}(\mathbf{u}_1,\mathbf{u}_1) &=&
                                            Q(\mathbf{V}^{(1)}
                                            ,\mathbf{V}^{(1)})a^2 P_{i_c}^2 + 2
                                            Q(\mathbf{V}^{(1)}, \bar{\mathbf{V}}^{(1)})\bar{a}a
                                            P_{i_c}\bar{P}_{j_c} + \nonumber \\
                                        & & Q(\bar{\mathbf{V}}^{(1)}, \bar{\mathbf{V}}^{(1)})
                                            \bar{a}^2 \bar{P}_{j_c}^2 . 
\end{eqnarray}

A key step is to expand the powers and products of $P_{i_c}$ using the
same basis functions, that is \cite{dougall}:
\begin{equation}
\label{eqn:rosapp}
P_{q} ^2 = \sum_{n=0}^{2 q} \xi_n P_n .
\end{equation}

Consider first $\mathbf{u}^*$ given by (\ref{eqn:uast1}), then
replacing (\ref{eqn:rosapp}) in (\ref{QU1U1}), and the result in
(\ref{FA1}), together with (\ref{solu1}), and applying the
orthogonality properties of the eigenfunctions of the Legendre
polynomials, we obtain

\begin{equation}
  \xi_{i_c} \langle \mathbf{V}^* \mid Q( \mathbf{V}^{(1)},
  \mathbf{V}^{(1)}) \rangle a^3 + p_1 \eta \langle  \mathbf{V}^*
  \mid J_p^c  \mathbf{V}_1 \rangle a^2-w_1\langle  \mathbf{V}^*
  \mid  \mathbf{V}_1 \rangle a \frac{\partial a}{\partial T}  =0 .
\end{equation}
In the case of odd Legendre polynomials, $\xi_{i_c} = 0$; hence, it is
sufficient to guarantee that the perturbation parameter and time scale
are canceled to ensure stable non-null patterns, that is, $p_1=0$ and
$w_1=0$.

Once the Fredholm-Alternative is fulfilled, for the solution of the
second order system, the following form is assumed:

\begin{align}
\mathbf{u}_2=& \sum_{s=0}^{2 i_c} \left( \mathbf{V}_{s}^{(2)} a^2
P_{s} + \bar{\mathbf{V}}_{s}^{(2)} \bar{a}^2
\bar{P}_{s}\right) + \mathbf{V}_{ij} a \bar{a}
P_{ic} \bar{P}_{jc} ,
\label{solu2app}
\end{align}
where $\mathbf{V}_{ij} = \mathbf{V}_{ij} (x,y)$.

Substituting (\ref{solu2app}) and (\ref{solu1}) into
(\ref{sisMulScala2}), and collecting the terms, we obtain the
following coefficient vectors:

\begin{align}
\mathbf{V}_{s}^{(2)}&=(-\eta J + k_{s0} D)^{-1}
                      \xi_{s} Q(\mathbf{V}^{(1)},\mathbf{V}^{(1)}) ,\\ 
\bar{\mathbf{V}}_{s}^{(2)}&= (-\eta J + k_{0s} D)^{-1}
                            \xi_{s}
                            Q(\bar{\mathbf{V}}^{(1)},\bar{\mathbf{V}}^{(1)})
  ,\\ 
\mathbf{V}_{ij}&= 2(-\eta J + k_{ij} D)^{-1} Q
                 (\mathbf{V}^{(1)},\bar{\mathbf{V}}^{(1)}) .
\end{align}

\subsubsection*{$O(\epsilon^3)$ equations}
We again apply the Fredholm Alternative as follows:

\begin{align}
    \label{FA3}
  \left\langle \mathbf{u}^* \middle\vert\ Q
  (\mathbf{u}_1,\mathbf{u}_2)+ C (\mathbf{u}_1, \mathbf{u}_1,
  \mathbf{u}_1) + p_2 \eta J_p^c  \mathbf{u}_1-w_2
  \frac{\partial \mathbf{u}_1}{\partial
  T}  \right\rangle =0 .
\end{align}

By using (\ref{terminocuadratico}) and (\ref{terminocubico}), and
(\ref{solu2app}) and (\ref{solu1}), the quadratic and cubic terms of
(\ref{FA3}) can be obtained:
\begin{eqnarray}
  \label{cuadratico}
  Q (\mathbf{u}_1,\mathbf{u}_2)&=& \sum_{s=0}^{2 i_c} \left ( Q(
                                         \mathbf{V}^{(1)},
                                         \mathbf{V}_s^{(2)})a^3 P_{i_c} P_s+
                                         Q( \mathbf{V}^{(1)},\bar{
                                         \mathbf{V}}_s^{(2)})a \bar{a}^2
                                         P_{i_c} \bar{P}_s+ \right . \nonumber \\
 & & Q(\bar{ \mathbf{V}}^{(1)}, \mathbf{V}_s^{(2)})\bar{a}a^2 \bar{P}_{j_c} P_s+
    Q(\bar{ \mathbf{V}}^{(1)},\bar{ \mathbf{V}}_s^{(2)})\bar{a}^3 \bar{P}_{j_c} \bar{P}_s+\nonumber\\
 & & \left . Q( \mathbf{V}^{(1)}, \mathbf{V}_{ij})a^2 \bar{a} P_{i_c}^2 \bar{P}_{j_c}+
                                         Q(\bar{ \mathbf{V}}^{(1)},
     \mathbf{V}_{ij})a \bar{a}^2 P_{i_c} \bar{P}_{j_c}^2 \right) ,
\end{eqnarray}

and

\begin{eqnarray}
  \label{cubico}
  C (\mathbf{u}_1,\mathbf{u}_1,\mathbf{u}_1) &=& C(
  \mathbf{V}^{(1)},
  \mathbf{V}^{(1)})a^3
  P_{i_c}^3 + 3C(
  \mathbf{V}^{(1)},\bar{
    \mathbf{V}}^{(1)})a^2\bar{a}P_{i_c}^2
  \bar{P}_{j_c}+ \nonumber \\
  & & 3C(\bar{ \mathbf{V}}^{(1)}, \mathbf{V}^{(1)})a
  \bar{a}^2 P_{i_c} \bar{P}_{j_c}^2 +
  C(\bar{ \mathbf{V}}^{(1)}, \bar{ \mathbf{V}}^{(1)}) \bar{a}^3\bar{P}_{j_c}^3 .
\end{eqnarray}
Once again the product of the eigenfunctions can be represented as an
expansion of the same eigenfunctions:

\begin{equation}
  \label{products}
  P_q P_{i_c} =\sum_{n}^{q+i_c} \zeta_n^{(q)} P_n , \;\;\;\;\; 
  P_{i_c}^3=\sum_{n}^{3 i_c} \chi_n P_n ,
\end{equation}
where the superscript $(q)$ in the first expression indicates that the
Fourier coefficient depends on the $q$-th eigenfunction.

Thus, substituting (\ref{products}) and (\ref{eqn:rosapp}) in
(\ref{cuadratico}) and (\ref{cubico}), we obtain

\begin{eqnarray}
  \label{cuadratico2}
  Q (\mathbf{u}_1,\mathbf{u}_2)&=&
                                         \sum_{s=0}^{2 i_c}\sum_{n=0}^{s+i_c}
                                          \left( Q(
                                         \mathbf{V}^{(1)},
                                         \mathbf{V}_s^{(2)})
                                           \zeta_n^{(s)}  a^3 P_n+
                                    Q(\mathbf{V}^{(1)},\bar{
      \mathbf{V}}_s^{(2)}) a \bar{a}^2 P_{i_c}
                                   \bar{P}_s \right .+ \nonumber \\
  & &  Q(\bar{ \mathbf{V}}^{(1)}, \mathbf{V}_s^{(2)})\bar{a}a^2
                                         \bar{P}_{j_c} P_s + 
                                   Q(\bar{
                                           \mathbf{V}}^{(1)},\bar{
                                           \mathbf{V}}_s^{(2)})
                                           \zeta_n^{(s)} 
                                         \bar{a}^3 \bar{P}_n+ \nonumber \\
  & & \left . Q( \mathbf{V}^{(1)}, \mathbf{V}_{ij}) \xi_s a^2
      \bar{a} P_s \bar{P}_{j_c}+  Q(\bar{ \mathbf{V}}^{(1)},
      \mathbf{V}_{ij}) \xi_s  a \bar{a}^2 P_{i_c}
                                         \bar{P}_s \right ),
\end{eqnarray}

and

\begin{eqnarray}
  \label{cubico2} 
  C (\mathbf{u}_1, \mathbf{u}_1, \mathbf{u}_1) &=& \sum_{n=0}^{3 i_c} \left( C( \mathbf{V}^{(1)},
                              \mathbf{V}^{(1)}) \chi_{n}
                              a^3 P_n + 3 C(
                              \mathbf{V}^{(1)}, \bar{ \mathbf{V}}^{(1)})
                              \xi_{n}  a^2 \bar{a} P_n \bar{P}_{j_c}
                                                   \right . +
                              \nonumber \\
                          & & \left . 3 C(\bar{ \mathbf{V}}^{(1)},
                              \mathbf{V}^{(1)}) \xi_{n}  a \bar{a}^2
                              P_{i_c} \bar{P}_n+  C(\bar{ \mathbf{V}}^{(1)},\bar{
                              \mathbf{V}}^{(1)}) \chi_{n}
                              \bar{a}^3\bar{P}_n  \right ) .
\end{eqnarray}

Now, (\ref{cuadratico2}), (\ref{cubico2}) and $\mathbf{u}^*$ given by
(\ref{eqn:uast1}) are replaced into (\ref{FA3}), and taking into
account that $\zeta_{i_c}^{(q)} =0$ for $q\leq i_c$ we get

\begin{align}
  \label{EqAmplitud0}
  \left\langle  \mathbf{V}^* \middle\vert\ \sum_{s=i_c+1}^{2 i_c} Q( \mathbf{V}^{(1)},
  \mathbf{V}_s^{(2)}) \zeta_{i_c}^{(s)} \right\rangle a^4 \delta_{i0}+
\left\langle  \mathbf{V}^* \middle\vert\ Q( \mathbf{V}^{(1)}, \bar{
  \mathbf{V}}_0^{(2)}) \right\rangle a^2 \bar{a}^2 \delta_{i0} 
  +  \nonumber\\
  \left\langle  \mathbf{V}^* \middle\vert\ Q(\bar{ \mathbf{V}}^{(1)},
  \mathbf{V}_{ij}) \xi_0 \right\rangle a^2\bar{a}^2 \delta_{i0} +
  \left\langle  \mathbf{V}^* \middle\vert\  C( \mathbf{V}^{(1)},
  \mathbf{V}^{(1)}) \chi_{i_c} \right\rangle a^4
  \delta_{i0}+ \nonumber \\
  \left\langle  \mathbf{V}^* \middle\vert\ 3 C(\bar{ \mathbf{V}}^{(1)},
  \mathbf{V}^{(1)}) \xi_{0} \right\rangle a^2 \bar{a}^2 \delta_{i0}+ 
  \left\langle  \mathbf{V}^* \middle\vert\ p_2 \eta J_p^c  \mathbf{V}^{(1)} \right\rangle a^2
  \delta_{i0} - \nonumber\\
  \left\langle  \mathbf{V}^* \middle\vert\
  \mathbf{V}^{(1)}\right\rangle a \frac{d a}{dT} \delta_{i0}=0 ,
\end{align}
where $\delta_{i0}=\int_{\Omega} (P_i \bar{P}_0)^2 dxdy$.

Equation \ref{EqAmplitud0} can be recasted as

\begin{align}
  \frac{1}{2} \left\langle  \mathbf{V}^* \middle\vert\ \mathbf{V}^{(1)}
  \right\rangle  \frac{d \lvert a \rvert ^2}{dT} 
  = \left\langle  \mathbf{V}^* \middle\vert\ \sum_{s=i_c+1}^{2 i_c} Q( \mathbf{V}^{(1)},
  \mathbf{V}_s^{(2)}) \zeta_{i_c}^{(s)} + C( \mathbf{V}^{(1)},
  \mathbf{V}^{(1)}) \chi_{i_c} \right\rangle  a^4 +\nonumber\\
  \left\langle  \mathbf{V}^* \middle\vert\ Q(\bar{ \mathbf{V}}^{(1)},
  \mathbf{V}_{ij}) \xi_0 + Q( \mathbf{V}^{(1)},\bar{
  \mathbf{V}}_0^{(2)}) + 3 C(\bar{ \mathbf{V}}^{(1)},
  \mathbf{V}^{(1)}) \xi_{0} \right\rangle_w  a^2\bar{a}^2+ \nonumber \\
  \left\langle  \mathbf{V}^* \middle\vert\  p_2 \eta J_p^c  \mathbf{V}^{(1)}\right\rangle a^2 .
\label{EqAmplitudfin}
\end{align}

Finally, by repeating the same procedure, but for the second null
vector (\ref{eqn:uast2}), the Stuart-Landau amplitude equations are
obtained:

\begin{align}
\frac{d \lvert a \rvert ^2}{dT}&=\alpha  \lvert a \rvert ^4+\beta
 \lvert a \rvert ^2 \lvert \bar{a} \rvert ^2+\theta  \lvert a \rvert ^2 ,\nonumber\\ 
\frac{d  \lvert \bar{a} \rvert ^2}{dT}&=\alpha  \lvert \bar{a} \rvert ^4+\beta
 \lvert a \rvert ^2  \lvert \bar{a} \rvert ^2+\theta
 \lvert \bar{a} \rvert ^2 ,
\label{Eq:Landauapp}
\end{align}

where 
\begin{align}
E&= \frac{1}{2} \left\langle  \textbf{V}^* \middle\vert\ \textbf{V}^{(1)}
   \right\rangle , \nonumber\\ 
 \alpha&= \frac{1}{E}\left\langle  \textbf{V}^* \middle\vert\ \sum_{s=i_c+1}^{2 i_c} Q(
         \textbf{V}^{(1)}, \textbf{V}_s^{(2)}) \zeta_{i_c}^{(s)}  + (
         \textbf{V}^{(1)}, \textbf{V}^{(1)})\chi_{i_c} \right\rangle ,  \\
\beta &= \frac{1}{E} \left\langle  \textbf{V}^* \middle\vert\ Q(\bar{
        \textbf{V}}^{(1)}, \textbf{V}_{ij}) \xi_0 + Q(
        \textbf{V}^{(1)}, \bar{ \textbf{V}}_0^{(2)}) + 3 C( \bar{
        \textbf{V}}^{(1)}, \textbf{V}^{(1)}) \xi_{0} \right\rangle , \\
\theta &= \frac{1}{E} \left\langle  \textbf{V}^* \middle\vert\ p_2
         \eta J_p^c  \textbf{V}^{(1)} \right\rangle .
\end{align}

\subsection{Stability of the Stuart-Landau equations}
The system (\ref{Eq:Landauapp}) has 4 equilibrium points:

\begin{align}
  &1)\hspace{0.5cm}\lvert a \rvert ^2=0,  \hspace{1cm}\lvert \bar{a}
    \rvert ^2=0\\ 
  &2)\hspace{0.5cm}\lvert a \rvert ^2=0,  \hspace{1cm}\lvert \bar{a}
    \rvert ^2=-\frac{\theta}{\alpha}\\ 
  &3)\hspace{0.5cm}\lvert a \rvert ^2=-\frac{\theta}{\alpha},
    \hspace{1cm}\lvert \bar{a} \rvert ^2=0\\ 
  &4)\hspace{0.5cm}\lvert a \rvert ^2=-\frac{\theta}{\alpha+\beta},
    \hspace{1cm}\lvert \bar{a} \rvert ^2=-\frac{\theta}{\alpha+\beta} 
\end{align}

We are interested in the parameter regions where the equilibrium
points are stable so that the Turing patterns are maintained for a
long time. Using the corresponding Jacobian matrices, the conditions
for the stability of each equilibrium point can be established to
obtain the results summarized in Table \ref{tab:LandauSolution}.

\begin{table}[h]
  \begin{center}
      \caption{Steady states and conditions for linear stabilities of the amplitude functions.}
		\label{tab:LandauSolution}
		\begin{tabular}{c | c | c }
                  \hline
                  Steady state &Conditions for linear stability &
                                                                  spatial pattern\\ 
                  \hline
                  $\lvert a\rvert ^2=\lvert \bar{a}\rvert ^2=0$ & $\theta<0$ & None \\
                  $\lvert a\rvert ^2=0,\lvert \bar{a}\rvert ^2=\frac{-\theta}{\alpha}$ &
                                                                                         $\theta>0$ and $\frac{\beta}{\alpha}>1$ & Stripes \\ 
                  $\lvert a\rvert ^2=\frac{-\theta}{\alpha},\lvert \bar{a}\rvert ^2=0$ &
                                                                                         $\theta>0$ and $\frac{\beta}{\alpha}>1$ & Stripes \\
                  $\lvert a\rvert ^2=\lvert \bar{a}\rvert ^2=\frac{\theta}{\alpha+\beta}$&
                                                                                           $\theta>0$ and $\alpha<-\lvert \beta\rvert<0$ & Spots \\
                  \hline
		\end{tabular}
	\end{center}
\end{table}

\section{Numerical simulations}
 \label{sec:numerical}

Here we consider the two-dimensional Schnakenberg system
(\ref{eqn:schnakenberg2d}) with zero-flux boundary conditions
(\ref{eqn:boundaries}). For this system, the results of the non linear
analysis, summarized in Table \ref{tab:LandauSolution} can be used to
divide the parameter space $(a,b)$ into domains corresponding to
different spatial patterns. For comparison purposes, homogeneous
diffusion, $\mathcal{D}(x) = 1$, and Legendre diffusion, where
$\mathcal{D}(\mathbf{x})$ is given by (\ref{SturmModel}), are
considered.

\begin{figure}[t!]
	\centering
	\begin{tabular}{c}
		\includegraphics[width=12cm]{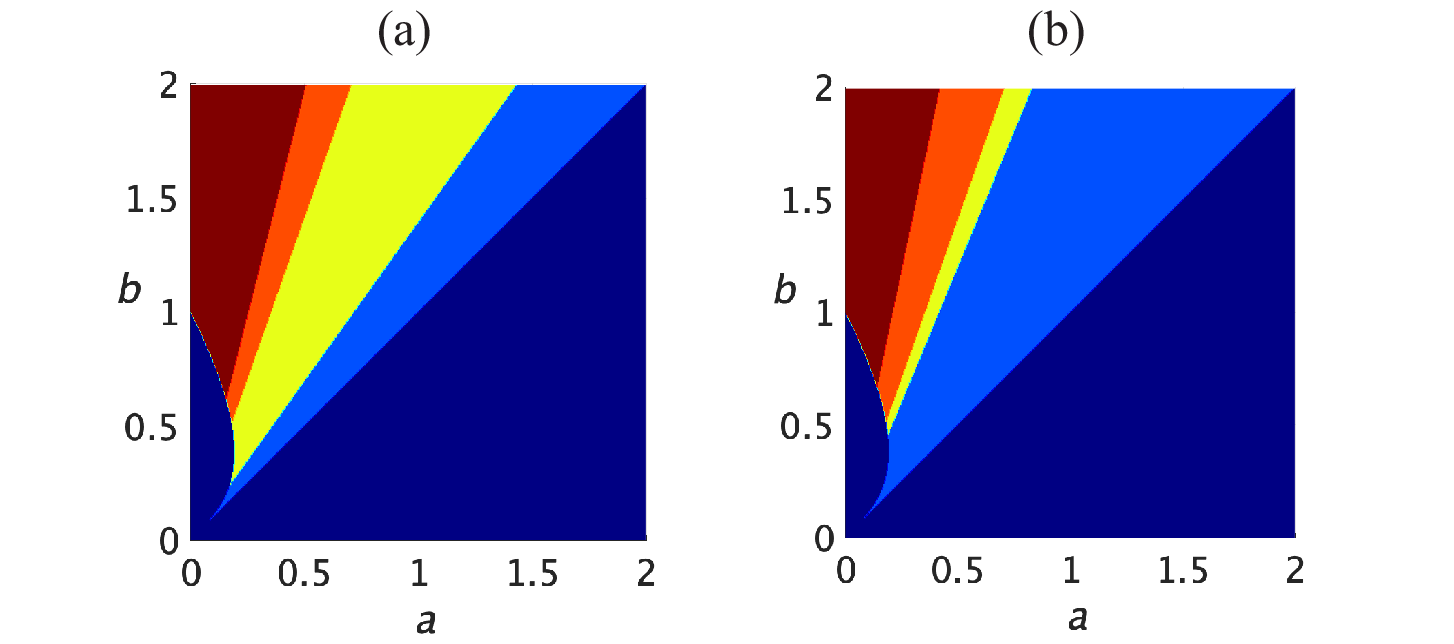}
	\end{tabular} 
	\caption{Parameter space $(a,b)$ of the Schnakenberg model
          with homogeneous diffusion (a) and Legendre diffusion
          (b). Red color represents stripes, orange color represents
          spots, yellow color represents patterns that cannot be
          predicted by the nonlinear analysis, and blue light and dark
          colors represent the region where the stability and
          instability Turing conditions are not fulfilled,
          respectively.}
	\label{Fig:DominioDeParametros}
\end{figure}
The results are shown in Fig. \ref{Fig:DominioDeParametros}, where the
regions where spatial patterns are stripes, spots, when the non linear
analysis cannot predict a particular pattern and when the conditions
for a Turing instability are not satisfied are indicated with colored
regions. Red color represents stripes, orange color represents spots,
yellow color represents patterns that cannot be predicted by the non
linear analysis, and blue light and dark colors represent the region
where the stability and instability Turing conditions are not
fulfilled, respectively. Fig. \ref{Fig:DominioDeParametros}a
corresponds to the case of homogeneous diffusion, where a spatial mode
$k_c = 4 \pi$, that is, $n=4$, was
chosen. Fig. \ref{Fig:DominioDeParametros}b corresponds to the case of
Legendre diffusion with spatial modes $k_c=3$, which corresponds to
$k_{30}= k_{03}= 12$ in (\ref{eqn:lind2}). The procedure for
generating the plots is as follows. For each of the two cases, with
the established value of $k_c$, for a given mapped point $(a,b)$, a
suitable value of $\eta$ is calculated using (\ref{eqn:lcS}) and $d_c$
is calculated using (\ref{eqn:dcS}). With these values, the parameters
$\alpha$, $\beta$ and $\theta$ were evaluated, and the conditions
listed in the second column of Table \ref{tab:LandauSolution} were
used to determine the type of pattern that was generated. The regions
in red are for patterns that cannot be predicted by the present
nonlinear analysis, which means that for $(a, b)$ in these regions
none of the four conditions for linear stability of the Stuart-Landau
equations, in the second column of Table \ref{tab:LandauSolution}, is
satisfied. However, because for these values of $(a,b)$ the conditions
for Turing instability are fulfilled, a pattern is formed. This
pattern is tipically observed with an irregular distribution of spots.

It is observed that the region of patterns that cannot be predicted by
the linear analysis is reduced in the case of Legendre diffusion as
compared with homogeneous diffusion. By contrast, the region where the
Turing stability condition is not fulfilled is enlarged.

The two-dimensional Schnakenberg system is solved to numerically
verifythe predictions for Legendre diffusion in
Fig. \ref{Fig:DominioDeParametros}b for some values of parameters $a$
and $b$. Regarding homogeneous diffusion,
Fig. \ref{Fig:DominioDeParametros}a was already obtained for the
Schnakenberg system in \cite{zhu} (Fig. 8b in that work), and the
plots match.

\begin{figure}[t!]
	\centering
	\begin{tabular}{c}
          \includegraphics[width=10cm]{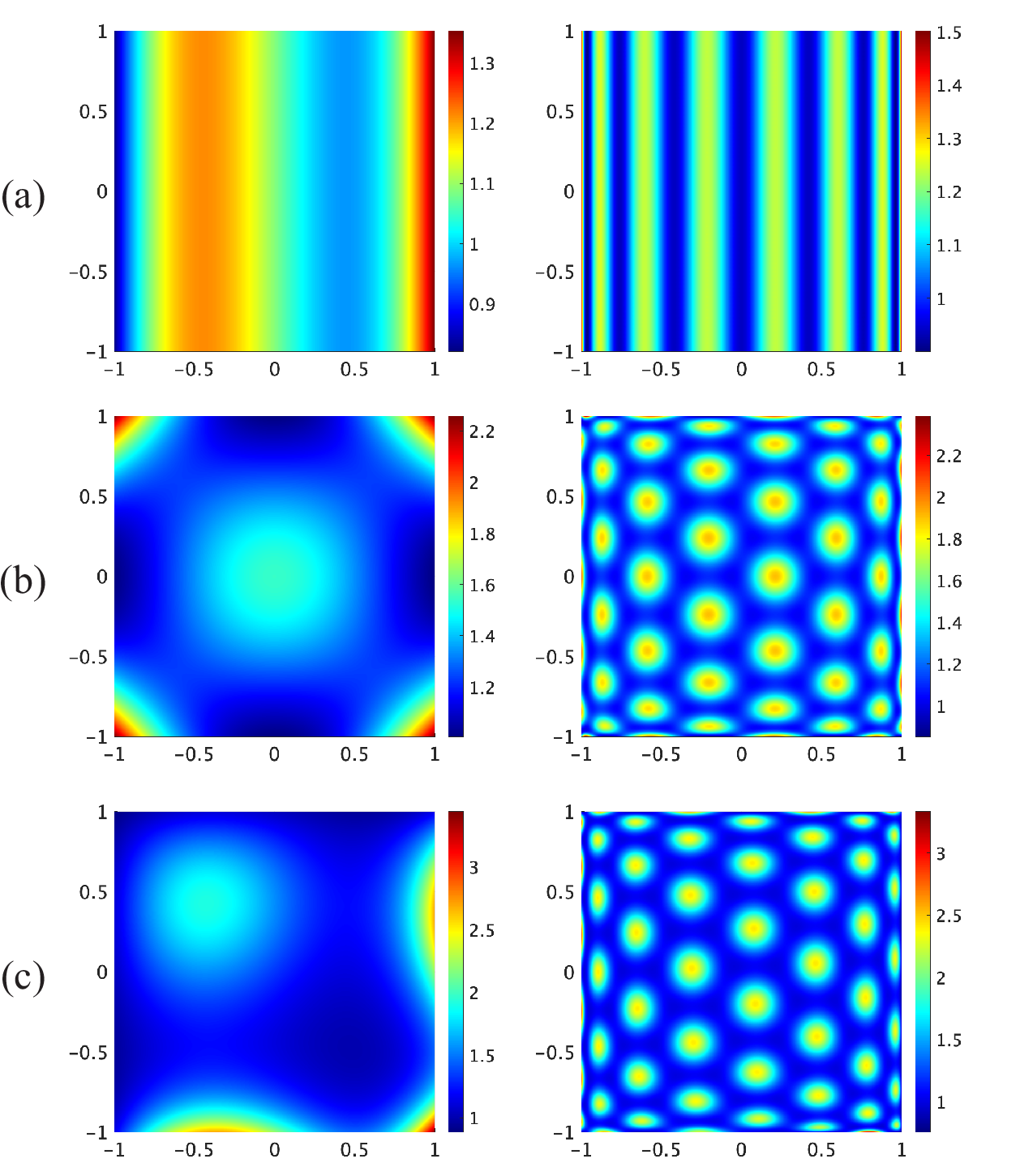}
	\end{tabular}
	\caption{Two-dimensional patterns obtained numerically for
          three sets of parameters $(a,b)$, and two values of $k_c$.
          The left plots correspond to $k_c=3$ and the right plots
          correspond to $k_c=14$. The three rows correspond to
          different values of $(a,b)$: \textbf{(a)} $(0.1, 1.0)$ in
          the region of stripes, \textbf{(b)} $(0.3, 1.0)$ in the
          region of spots, and \textbf{(c)} $(0.375, 1.0)$ in the
          region where our linear analysis cannot predict the type of
          pattern; in all cases $u$ is plotted. The parameter values
          are presented in Table \ref{tab:parameters}.}
	\label{fig:NumericalTest} 
\end{figure}

Because the nonlinear analysis is performed at critical values, once a
pair $(a,b)$ is selected, $d$ is chosen to satisfy (\ref{eqn:dcS}) and
$\eta$ is obtained from (\ref{eqn:lcS}) or (\ref{eqn:lcS2}) to satisfy
a given $k_c$. To generate an initial growth of the spatial pattern,
parameters $a$ and $b$ were perturbed by $\Delta a = 0.001$ and
$\Delta b = 0.01$.

Three sets of parameters $(a,b)$ were chosen in different regions of
Fig. \ref{Fig:DominioDeParametros}b, and they are given in Table
\ref{tab:parameters}. Two cases were considered, $k_c = 3$ (as in
Fig. \ref{Fig:DominioDeParametros}b) and $k_c = 14$.

\begin{table}[t!]
  \begin{center}
      \caption{The model parameters of the Schnakenberg system
        (\ref{eqn:schnakenberg2d}) used in
        Fig. \ref{fig:NumericalTest}. $(a,b)$ were chosen at different
        regions in Fig. \ref{Fig:DominioDeParametros}b. For each
        $(a,b)$, $d_c$ was calculated using (\ref{eqn:dcS}). Two
        values of $k_c$ are used: $k_c = 3$ and $k_c = 14$. The
        corresponding values of $\eta$ were obtained using
        (\ref{eqn:lcS2}).} 
		\label{tab:parameters}
                \begin{tabular}{|c|l|c|c|c|}
\hline 
$(a,b)$ & Region & $d_c$ & $k_c$ & $\eta$ \\
\hline 
\multirow{2}{*}{$(0.1,1.0)$} & \multirow{2}{*}{Stripes} &
                                                          \multirow{2}{*}{0.1003}
                         & 3 & 3.4552 \\
 &  &  & 14 & 60.466\tabularnewline
\hline 
\multirow{2}{*}{$(0.3,1.0)$} & \multirow{2}{*}{Spots} & \multirow{2}{*}{0.0342} & 3 & 1.7066\tabularnewline
 &  &  & 14 & 29.866\tabularnewline
\hline 
\multirow{2}{*}{$(0.375,1.0)$} & \multirow{2}{*}{Not predicted} & \multirow{2}{*}{0.0224} & 3 & 1.3078\tabularnewline
 &  &  & 14 & 22.866\tabularnewline
\hline 
\end{tabular}

	\end{center}
\end{table}

The resulting patterns are shown in Fig. \ref{fig:NumericalTest}. We
observed that the predictions of the nonlinear analysis were numerically
validated. In addition, the spatial variation in the diffusion rate, in
accordance with Fig. \ref{fig:figure2}, is clearly observed in the
figures on the right.

\section{Discussion}

A problem of pattern formation with reaction-diffusion equations with
space varying diffusion is discussed. This problem corresponds to
diffusion coefficients with explicit dependence on spatial
variables. The first part of the mathematical analysis of this
problem, the linear analysis, can be performed for general
space-dependent diffusion coefficients \cite{vangorder}, and the main
results are reviewed in Sec. \ref{sec:linear}. The second part
however, requires a generalization of the standard weakly nonlinear
analysis for eigenfunctions other than the eigenfunctions of the
Laplace operator. In this study, we consider the particular case of
the operator $\partial _x \left( (1-x^2) \partial _x \right)$, whose
eigenfunctions are the Legendre polynomials $P_n(x)$, and eigenvalues
$n (n+1)$. For the two-dimensional problem we propose a tensor form
for the spatial variation of the diffusion rate, which leads to a
Sturm-Liouville problem with eigenfunctions given by
$P_{ij}(x,y) = P_i(x) P_j (y)$ and eigenvalues $i(i+1) + j(j+1)$. With
these functions, a generalization of the weakly nonlinear analysis is
proposed. This generalization allows us to find conditions for the
formation of stripped or spotted patterns, which are verified
numerically, and compared with the case of homogeneous diffusion,
using the Schnakenberg reaction-diffusion system as an example .

Some differences arise with the standard non linear analysis, of which
it is worth highlighting the appearance of Fourier coefficients
$\xi_n$, $\zeta_n ^{(q)}$, and $\chi_n$, which vary with the critical wave
number, contrary to the case of the eigenfunctions of the Laplace
operator where they remain constant. This can produce variations in
terms of the amplitude equations and asymmetries when the critical
wavenumber is modified, which is worth exploring further.

It must be said that it is tempting to attempt a generalization of
nonlinear analysis for orthogonal eigenfunctions of any
Sturm-Liouville problem, where the Legendre functions are a particular
case. This general approach faces several difficulties, mainly with
the Fredholm alternative, which we are currently working on.

The type of patterns generated by the particular problem studied in
this work enriches the field of pattern formation and the
generalization of the nonlinear analys developed here can also be of
interest in other fields such as climate modeling \cite{north,hetzer}
and can motivate further generalization by using orthogonal
eigenfunctions of any Sturm-Liouville problem.

\section*{Acknowledgments}

This work was supported by CONACYT under Project No. A1-S-8317. Elkinn
A. Calder\'{o}n-Barreto received Fellowship 995896 from CONACYT.


\end{document}